\documentclass[prb,twocolumn,amsmath,amssymb,superscriptaddress]{revtex4-2}
\usepackage[T1]{fontenc}
\usepackage{customcommands}
\usepackage{amsmath}
\usepackage{amsthm}
\usepackage{amsfonts}
\usepackage{xcolor}
\usepackage{slashed}
\usepackage{mathtools}

\def\eq#1 { \begin{equation} #1 \end{equation} }
\def\eqn#1{ \begin{eqnarray} #1 \end{eqnarray} }
\def\nn { \nonumber }

\def\ba{\vec{a}}

\def\bQ{\vec{Q}}

\def\bl{\vec{\ell}}

\def\bx{\vec{x}}
\def\by{\vec{y}}

\def\bpartial{\vec{\partial}}

\begin{document}

%
%

%
%
\title{Curved-space Dirac description of elastically deformed monolayer graphene is generally incorrect}

\author{Matthew M.~Roberts}
\email{matthew.roberts@imperial.ac.uk}

\author{Toby Wiseman}
\email{t.wiseman@imperial.ac.uk}

\affiliation{Theoretical Physics Group, Blackett Laboratory, Imperial College, London SW7 2AZ, United Kingdom}


\begin{abstract}

Undistorted monolayer graphene has energy bands which cross at protected Dirac points. 
It elastically deforms and much research has assumed the Dirac description persists, now in a curved space and coupled to a gauge field related to lattice strain. 
We show this is incorrect by using a real space gradient expansion to study how the Dirac equation derives from the tight binding model. Generic spatially varying hopping functions give rise to large magnetic fields which spoil the truncation in derivatives.
In the perturbative regime, the only consistent truncation to Dirac is one with nontrivial gauge field but in flat space.
One can instead fine tune the magnetic field to be small, and we derive the resulting differential condition that the hopping functions  must satisfy to yield a consistent truncation to Dirac in curved space. We consider whether mechanical effects might impose this fine tuning, but find this is not the case for a simple elastic membrane model.

\end{abstract} 

\maketitle


\section{Introduction}

Monolayer graphene has a band structure which contains two protected massless Dirac cones at the $K$ and $K'$ points. When undoped the bands are at half filling, putting the chemical potential at the Dirac points  \cite{novoselov2005two,novoselov2005two,Zhang:2005zz}. 
It can be bent beyond the linear regime \cite{lee2008measurement}, and when freely 
 suspended, it naturally ripples~\cite{meyer2007structure,fasolino2007intrinsic}. 
When considering transport in such systems it is imperative to derive a continuum description of such distorted lattices. 

Here we focus on the tight-binding model for graphene. For perturbative distortions Fourier space calculations appear to show an effective description of Dirac fields in curved space coupled to a gauge field proportional to the strain of the lattice, originally for carbon nanotubes \cite{sasaki2005local} and later for graphene \cite{dejuan2012spacedepfermi,Zubkov:2013sja,oliva2015generalizing,yang2015dirac,si2016strain,khaidukov2016landau,oliva2017low,wagner2019quantum}\footnote{Work predating the discovery of graphene focused on continuum descriptions of fullerines \cite{gonzalez1992continuum,gonzalez1993electronic}, as well as topological lattice defects \cite{cortijo2007effects,cortijo2007electronic}, neither of which which we are considering here. For a modern review see \cite{vozmediano:physrev}}. 
It was noted early on that the magnetic field of this ``strain gauge field'' scales inversely with the lattice spacing, $\vec{A} \sim \mathcal{O}(1/a)$.
This is naively concerning as it appears the magnetic fields induced by strain and curvature would be very large. However it is suppressed by the perturbative expansion.
Based on these analyses much work has assumed a curved space Dirac description exists \cite{de2007charge,guinea2008gauge,vozmediano2008gauge,de2013gauge,iorio2015revisiting,arias2015gauge,stegmann2016current,castro2017pseudomagnetic,Golkar:2014paa,Golkar:2014wwa}.
Our main result is to show that actually this inverse scaling of the magnetic field with the lattice spacing in fact does obstruct such a continuum Dirac description in curved space.

Here we derive an effective low energy description from the tight binding model via a real space gradient expansion, where we assume the hopping strengths vary on scales much larger than the lattice spacing. We consider both the leading Dirac term and the next two derivative correction.
We argue that the previous perturbative Dirac descriptions are inconsistent in that, while a correction to the frame adding curvature exists, it is of the same order as the higher derivative term due to the magnetic field scaling as $1/a$. Hence in the perturbative case the correct gauged Dirac description is one in flat space. Worse still, in the non-linear regime we argue that all higher derivative terms are of the same order and so there is no truncation whatsoever.
In order to consistently keep curvature  the hopping strengths must be fine tuned to remove large magnetic fields. When this unnatural fine tuning is made, we  derive a gauged Dirac description non-linearly in the hopping functions, which lives in a curved space with torsion free spin connection.

\section{The lattice model and the continuum}
\label{sec:continuum}

 The tight-binding Hamiltonian for graphene is
\eqn{
\label{eq:gen_hopping_graphene_position}
H &=&\sum_{n,\bx_A} \left(
t_{n,\bx_A+\frac{a\bl_n}{2}} a^\dagger_{\bx_A} b_{\bx_A +a \bl_n} + \mathrm{h.c.}\right)
}
where $t_{n,\bx}$ is the (real valued) hopping strength in one of the lattice translation directions $n$, $a_{\bx_A}^\dagger,~b_{\bx_B}^\dagger$ are creation operators on the respective sublattices $A$ and $B$ and $a\bl_n$ are the translations from vertices in $A$ to its neighbours, $\bl_1 =  (\sqrt{3}/2,1/2),~\bl_2 =  (-\sqrt{3}/2,1/2),~\bl_3  = -\bl_1 - \bl_2$. Note we have put in an explicit lattice spacing $a$ (it is implicitly in the lattice coordinates as e.g. $\bx_A = a \left(m_A\ba_1 + n_A \ba_2 \right),~\ba_{1,2} = \bl_{1,2} - \bl_3$). Note that the $t_n$ take values on the links, not the vertices. A general one-particle state is 
\eq{
\ket{\Psi(t)} = \left(  \sum_{\by_A} A_{\by_A}(t) a^\dagger_{\by_A}+ \sum_{\by_B} B_{\by_B}(t) b^\dagger_{\by_B} \right) \ket 0
}
and the Schr\"odinger equation  $ i \hbar \partial_t \ket \Psi = H\ket\Psi $ gives
\eqn{
\label{eq:lattice}
 i \hbar \partial_t  A_{\bx_A} = \sum_n t_{n, \bx_A +\frac{a \bl_n}{2}} B_{\bx_A +a \bl_n} ,\nn \\
 i \hbar \partial_t  B_{\bx_B} =  \sum_n t_{n, \bx_B -\frac{a \bl_n}{2}} A_{\bx_B -a \bl_n} .
}
Firstly we take the hopping parameters to slowly vary, so we may write $t_{n,\bx} = t_n(\bx)$ where $t_n(\bx)$ are smooth functions of the coordinates $x^i = (x,y)$. 
We may then think of a continuum limit as we refine the lattice taking $a \to 0$. We write the lattice wavefunctions $A, B$ in terms of slowly varying wavefunctions $F,G$ and a rapidly oscillating phase $\Phi(\bx)/a$, 
\eqn{\label{eq:first_field_redef}
A_{\bx}(t) &=& F(t,\bx) f(\bx) e^{i \left( \frac{\Phi(\bx)}{a} +  \frac{ \phi(\bx)}{2}\right)} \; , \nn \\
\quad B_{\bx}(t) &=& G(t,\bx) f(\bx) e^{i \left( \frac{\Phi(\bx)}{a} - \frac{ \phi(\bx)}{2}\right)}
}
where we assume the phases $\Phi$ and $\phi$, and rescaling function $f$ are smooth functions of $x^i$ and $a$, so smooth in the continuum limit $a \to 0$.
Finding a $\Phi(\vec{x})$ to eliminate oscillations in $\Psi(\vec{x})$ that diverge as $a \rightarrow 0$ means that it is unique up to $\mathcal{O}(a)$ corrections, which are naturally interpreted as gauge transformations and will come with a compensating transformation of the strain gauge field, as we will see below. The subleading correction $\phi(\vec{x})$, under which $F,G$ have \emph{opposite} charge, is similarly unfixed, and can naturally be interpreted as an $SO(2)$ rotation of the frame as we shall see shortly. As we are interested in low energy states and the lattice theory is static, the only time dependence is seen in the slowly varying modulation functions $F$ and $G$. 
However at this stage we will not assume these are smooth in the $a \to 0$ limit, only that we may perform a gradient expansion in $a \partial_i$. 
Now expanding in $a$ we may write the Schr\"odinger equation as the continuum equation,
\begin{widetext}
\eqn{
\label{eq:continuum}
 \frac{i\hbar}{T} \gamma^0 \partial_t \Psi &=&  \gamma^a \left( u_a + i a z_a  + a^2 r_a  \right) \Psi   - i  a  \gamma^a \left( w^i_a - i  a q^i_a   \right) \partial_i \Psi
  +  \frac{a^2}{2} \gamma^a v^{ij}_a \partial_i \partial_j \Psi + O(a^3)
}
\end{widetext}
where $\Psi = ( F, G )$ is a complex 
spinor and we choose Dirac matrices  $\gamma^A = (\gamma^0, \gamma^a) = ( - i \sigma^3, \sigma^1, \sigma^2 )$ where $\sigma^i$ are Pauli matrices and $\partial_i = \partial / \partial x^i$ are lattice coordinate derivatives. The quantities $u_a, z_a, r_a, w^i_a, q^i_a, v^{ij}_a$ are real and the ones we require here are, 
\eqn{
\label{eq:uwvz}
I^a u_a &=& \frac{2 ie^{ i \phi} }{3 T} \sum_n e^{- i  \partial_n {\Phi} }   t_n \, , \;
I^a w^i_a = \frac{2 e^{ i \phi} }{3 T} \sum_n \ell^i_n e^{- i  \partial_n {\Phi} } t_n \; ,\nn \\ 
I^a v^{ij}_a &=&  \frac{2 ie^{ i \phi} }{3 T} \sum_n \ell^i_n \ell^j_n e^{- i  \partial_n {\Phi} } t_n \, , \; 
I^a z_a = - \frac{f^2}{2} \partial_i \left( \frac{w^i}{f^2} \right)  
}
with $I^a = ( 1, i)$ and  $\partial_n =\vec{\ell}_n\cdot \vec{\partial}$. We introduce the energy scale $T$ and think of $t_n(\vec{x}) = T$ as the undeformed model.

\section{Attempting to truncate to Dirac}

We will now truncate this continuum Schr\"odinger equation to first derivative order acting on $\Psi$. We should be suspicious about whether such a truncation is valid, but for now we will interpret
\eqref{eq:continuum} to order $O(a)$ as a curved space Dirac equation coupled to a gauge field.
We compose spacetime coordinates $x^\mu = (t, x^i)$ and define the metric,
\eq{
\label{eq:spacetime}
ds^2 = g_{\mu\nu} dx^\mu dx^\nu = - v^2 dt^2 + g_{ij}(\vec{x}) dx^i dx^j \, .
}
Here $v$ will be the Fermi velocity for the undeformed model, $t_n(\vec{x}) = T$. 
This may be written in terms of a frame $e_A^\mu$, and its dual $e^A_\mu$, where,
\eq{
e_0^t = \frac{1}{v} \; , \quad e_a^t = e_0^i = 0 \; , \quad g_{\mu\nu} = \eta_{AB} e^A_\mu e^B_\nu
}
with $\eta_{AB} = \mathrm{diag}(-1,+1,+1)$.
We fix 
\eq{
\label{eq:f}
 f = \sqrt{| \det{e_i^a} |} = |g|^{1/4}
}
which ensures that the $U(1)$ charge density of the Dirac theory is that of the original electrons,
\eq{
J^0 =  \sqrt{g} \bar{\Psi} \gamma^0 \Psi =(A, B)^\dagger \!\cdot\! (A, B) \, .
}
This is equivalent to ensuring the lattice anti-commutators $\{a^\dagger_{\bx}, a_{\by}\} =\{b^\dagger_{\bx}, b_{\by}\} = \delta_{\bx,\by}$ imply the correct curved space anti-commutator $\{\Psi^\dagger(\bx), \Psi(\by)\}=\frac{1}{\sqrt{g}}\delta^{(2)}(\bx-\by)$.

Now the Schr\"odinger equation~\eqref{eq:continuum}, truncated to first derivatives on spinors, can be written as,
\eq{\label{eq:covariant_two_deriv}
 a i e_a^\mu \gamma^a D_\mu \Psi = O(a^2)
}
where we have taken $v = 3 a T/(2 \hbar)$, and the covariant derivative is given in terms of a magnetic gauge field $A_\mu = (0, A_i)$ and spin connection, parameterized here by the spatial 1-form $\Omega_i$,
\eqn{
D_t \Psi = \partial_t \Psi \; , \quad D_i \Psi &=& \left(  \partial_i  - i A_i +  \frac{i}{2} \sigma_3 \Omega_i  \right) \Psi 
}
and the frame and gauge field to this order $O(a)$ are,
\eqn{
\label{eq:eA}
e^i_a = w^i_a \; , \quad  A_i = - \frac{1}{a} e^a_i u_a \; .
}
Using the relations~\eqref{eq:uwvz} and our choice of $f$ we find,
\eqn{
\Omega^i &=&  \epsilon_{ab} e^j_a \partial_j e^i_b
}
which is precisely the \emph{torsion free} spin connection that follows from the frame $e^i_a$. While we might expect that in the absence of lattice defects torsion vanishes in a continuum description, it is pleasing to see this explicitly emerge.
 
We may understand the local freedom of shifting the phases of the two lattice fields $A$ and $B$ in equation~\eqref{eq:first_field_redef} as a 
local frame rotation freedom,
\eqn{
\label{eq:frame}
\phi \to \phi + \delta\phi  \;  \implies \; \left\{ 
\begin{array}{ccc}
e_a^i &\to&  e_a^i -  \delta\phi \, \epsilon_{ab} e_b^i \\
A^i &\to& A^i
\end{array}
\right.
}
for an infinitessimal $\delta\phi$, together with a local gauge transform on the vector $A_a = e^i_a A_i$,
\eqn{
\label{eq:gauge}
\Phi \to \Phi + \delta \Phi  \; \implies \; \left\{
\begin{array}{ccc}
A_a &\to& A_a - \frac{1}{a} \partial_a \delta \Phi  \\
e_a^i &\to&  e_a^i - v^{ij}_a \partial_j \delta \Phi
\end{array}
\right.
}
for infinitessimal $\delta \Phi$.

Now we ask whether this truncation to first derivatives can be generally valid. We argue that for generic $t_n$ it is not, as contributions from the standard Dirac kinetic terms $\gamma^a e_a^\mu (\nabla_\mu - i A_\mu) \Psi$,  come at the same order as those from higher derivative terms in equation~\eqref{eq:continuum} such as $\gamma^a v^{ij}_a \partial_i \partial_j \Psi$. This may be seen in two ways;
\begin{itemize}
\item The gauge field in equation~\eqref{eq:eA} goes as $A_i \sim 1/a$. The spinor $\Psi$ responding to this will then generally have variation on scales that vanish as $a \to 0$, hence ruining the gradient expansion. We will discuss this explicitly for perturbative deformations.
\item The local phase symmetry in~\eqref{eq:gauge} is a gauge transformation for $A_a$, but $e_a^i$ transforms too and this is inconsistent with its interpretation as a frame which should be invariant. 
\end{itemize}
On the latter issue there should be a continuum formulation of~\eqref{eq:continuum} written to manifest these local symmetries where the derivative expansion will be in covariant derivatives with respect to the gauge and frame symmetry. Consider a putative two derivative term to match that in~\eqref{eq:continuum}. It will have the form $a^2 \gamma^a v^{ij}_a \tilde{D}_i \tilde{D}_j \Psi$ where $\tilde{D}_i$ are covariant. Note that their gauge and spin connections need not be the same as those of the leading Dirac theory. However, by taking the gauge field to scale as $\sim 1/a$ then we see this term contains a contribution,
\eqn{
\label{eq:higherderiv}
\frac{a^2}{2} \gamma^a v^{ij}_a \tilde{D}_i \tilde{D}_j \Psi \supset - i a^2 \gamma^a v^{ij}_a B_i \partial_j \Psi
}
for some gauge connection $B_i$. Then \eqref{eq:eA} gains a new term,
$w^i_a = e^i_a + a v^{ij}_a B_j$,
and now if $B_i$ transforms as $v^{ij}_a B_j \to v^{ij}_a B_j - \frac{1}{a} v^{ij}_a \partial_j \delta \Phi$, then we indeed see that $e^i_a$ is invariant with $B_i$ accounting for the transformation of $w^i_a$. However, then the gauge fields mix the contribution of covariant derivative terms between the partial derivative orders. This naturally extends to all higher derivative terms, with higher powers of the connection cancelling the $a$ supression.

\section{Two truncations to Dirac}
  
In order to give a consistent truncation of our theory to the leading Dirac term we must tame the lattice scale gauge field by requiring $a A_i \to 0$ as $a \to 0$.
There are two approaches.

\subsection{Perturbative deformation}

The lattice scale of the gauge field has been previously emphasised in~\cite{Zubkov:2013sja}. In the derivation of curved space Dirac of~\cite{dejuan2012spacedepfermi} and following work this was addressed using a perturbative expansion where,
\eqn{
t_n = T \left( 1 + \epsilon \delta t_n \right)
}
with $| \epsilon | \ll 1$. At $\epsilon=0$ we return to the undeformed lattice Hamiltonian, and can use standard momentum space tools. With our conventions a Dirac cone sits at the $K$ point with wave vector $\vec{K} = \frac{1}{a}(\frac{4\pi}{3 \sqrt{3}},0)$. This means that a slowly varying continuum field $\Psi$ is related to lattice wavefunctions via $(A,B) = e^{i \vec{K}\cdot \vec{x}}\Psi$. Motivated by this we will choose $\Phi$ to be this transformation at leading order,
\eqn{
\Phi = a \vec{K}\cdot x + \epsilon \chi(\bx) = \frac{4 \pi}{3 \sqrt{3} } x + \epsilon \chi(\bx) \, .
}
The gauge field is then,
\eqn{
A_i = - \frac{\epsilon}{a} \left( \frac{2}{3} \sum_n \epsilon_{ij} \delta^j_n \delta t_n + \partial_i \chi \right)
}
and we see $\chi(\bx)$ parameterizes the gauge freedom. While this goes as $\sim 1/a$, the perturbative expansion in $\epsilon$ controls this. 
The geometry depends on this gauge, the frame leading to the metric,
\eqn{
g_{ij} = \delta_{ij} - \frac{4}{3} \epsilon \sum_n  \ell^i_n  \ell^j_n  \delta t_n +  \epsilon K^{ijk} \ce^{k\ell} \partial_\ell \chi
}
which has Ricci scalar curvature
\eqn{\label{eq:pert_ricci}
R = \frac{4}{3} \epsilon \sum_n \left( \delta^{ij} - \ell^i_n \ell^j_n \right) \partial_i \partial_j \delta t_n +  \epsilon  K^{ijk}\ce^{k\ell}\partial_i \partial_j \partial_\ell \chi \qquad
}
where we have defined $K^{ijk} = - \frac{4}{3} \sum_n \ell_n^i \ell_n^j \ell_n^k$. 
Taking the $K'$ Dirac point corresponds to taking $\Phi \to - \Phi$, $e_1^i \to - e_1^i$ and $A_i \to - A_i$, and the metric is invariant. For $\chi = 0$ these reproduce the results of~\cite{dejuan2012spacedepfermi} (when we consider the physical metric, rather than the Weyl rescaled one~\cite{stegmann2016current}). Previously these results have been taken to show the perturbatively deformed tight binding model is described by a curved space Dirac equation. However we explicitly see here the gauge freedom $\chi$ gives a physical contribution to the curvature. 
Noting that 2-d geometry is locally characterized by the Ricci scalar, in fact we can then choose any geometry, including a flat one, with an appropriate gauge choice $\chi$. The second concerning feature highlighted in~\cite{Zubkov:2013sja} is that while we have controlled $A_i \sim \epsilon/a$, we have done this at the cost of the spin connection being parametrically smaller, $\Omega_i \sim \epsilon$, as $a \to 0$. In \cite{Zubkov:2013sja} it is argued that the spin connection should be ignored, leaving only a frame and gauge field. We will demonstrate that for general perturbations it is inconsistent to ignore the spin connection but not the variation of the metric, as they are of the same order, and further
that the corrections to the frame and spin connection have the same order as the two derivative term.
Consider in some region of size $L$ the perturbative deformation to be $\phi = \chi = \delta t_3 = 0$ and $\delta t_1 = - \delta t_2 = \frac{\sqrt{3} x}{2 L}$. This yields $A_i = \left( 0, \frac{\epsilon}{a} \frac{x}{L} \right)$, a Landau gauge magnetic field $B = \frac{1}{\ell_B^2}$ with $\ell_B = \sqrt{a L/\epsilon}$ the magnetic length which diverges as $a \to 0$ but may be parametrically larger than the lattice scale. 
At leading order $O(\epsilon)$ and $a \ll L$ this may be solved by Landau levels. Here we focus on the lowest level wavefunction, taking $\Psi = \left(  0 , e^{- \frac{x^2}{2 \ell_B^2}} \right) $. Now we can evaluate the Dirac term on this leading solution and compare this to the two derivative term. Then at $O(\epsilon)$ we have $v^{ij}_a =  \frac{1}{2} e_a^k K^{kmi} \epsilon_{mj}$, and find,
\eqn{
i a e^i_a \gamma^a \nabla_i \Psi = - \frac{i a \epsilon}{4 L} \sigma^1 \Psi  \, , \; a^2 \gamma^a v^{ij}_a \partial_i \partial_j \Psi = -\frac{i a \epsilon}{2 L} \sigma^1 \Psi,  \qquad  
}
so both go as $\sim a \epsilon/L$. We emphasize that the Dirac term is non-zero here due to the varying frame at $O(\epsilon)$.
Thus this simple example demonstrates that the perturbative contribution to the frame in the Dirac term is the same order as higher derivative terms, and it is therefore inconsistent to consider it in isolation.

We note it \emph{is} consistent to truncate to Dirac if we ignore corrections to the frame as then for $a \ll L$ these terms are small corrections to the flat gauged Dirac term. 

\subsection{Fine tuning}

In order to preserve curvature we are forced to fine tune the gauge field so that $a A_i \to 0$ as $a \to 0$, which implies the condition $u_a \to 0$ in this limit.
The condition $u_a = 0$ is solved by,
\eq{\label{eq:dphi_eq}
{t}_1 = \frac{\sin\left[(\bl_2 - \bl_3)\!\cdot \!\bpartial\Phi \right]}{\sin\left[(\bl_1 - \bl_2)\!\cdot\! \bpartial\Phi \right]}{t}_3, ~ 
{t}_2 = \frac{\sin\left[(\bl_1 - \bl_3)\!\cdot \!\bpartial\Phi \right]}{\sin\left[(\bl_2 - \bl_1)\!\cdot\!\bpartial\Phi \right]}{t}_3,
}
so that ${\Phi}$ determines the ratios ${t}_{1,2}/{t}_3$. However we wish to specify couplings ${t}_n$. We can solve \eqref{eq:dphi_eq} for $Q_x = \partial_x \Phi$ and $Q_y = \partial_y \Phi$ independently as functions of ${t}_n$,\footnote{Reality of $\Phi$ requires ${t}_1 + {t}_2>{t}_3>0$ and permutations. This is the same condition as for the homogeneous model \cite{pereira2009tight}.}
Explicitly if we define,
\eq{\label{eq:XY_defn}
{X} = e^{\frac{\sqrt{3} i}{2} Q_x} \; , \quad {Y} = e^{\frac{{3} i}{2} Q_y}
}
then we find,
\eq{
\label{eq:XY}
{Y} = - \frac{{t}_3 {X}}{{t}_2 + {t}_1 {X}^2} \; , \quad {X}^4 +  \frac{{t}_1^2 + {t}_2^2 - {t}_3^3}{ {t}_1 {t}_2}  {X}^2 + 1 = 0 
}
where the two roots of the quadratic in $X^2$ yield the inequivalent Dirac points. 
In the case of homogeneous $t_n$ where we can Fourier transform and work in momentum space, $\bQ/a$ given by the two roots are the wave-vectors of the two Dirac points.

However given $\vec{Q}$ one can only integrate to find a phase $\Phi$ if the integrability condition $\partial_x Q_y = \partial_y Q_x$ holds. Using the explicit solution for $\vec{Q}$ above, we find this condition can be written neatly as the constraint that  the $t_n$ must obey,
\eq{
\label{eq:constraint}
\sum_n V^i_{~n} \frac{\partial_i {t}_n}{{t}_n} = 0, ~ V^i_{~m} =\sum_n ( K^{ijk}\ell_m^j \ell_n^k -\frac{1}{2} \ell^i_m ) {t}_n^2 \, .
}
We will physically interpet this constraint shortly.

An important solution to this is constant but unequal $t_n$\cite{vozmediano:physrev}, which corresponds to constant nonzero strain. As we show below, in this case the metric has changed but is still constant, and so there is no geometric curvature. This corresponds to a misalignment of the lattice and ambient space coordinates. Using the preserved translational symmetry one can still work in momentum space, finding that the location of the Dirac cone has moved.

Another special solution where the $t_n$ are spatially varying is when they are equal, so $t_n = t(\bx)$. This corresponds to a pure expansion lattice distortion, with vanishing strain.

Suppose the above condition \eqref{eq:constraint} holds so we may find a solution to \eqref{eq:dphi_eq} written as  ${\Phi} = f( {t}_{1,2}/{t}_3 )$. Then a second solution is given by ${\Phi} = - f( {t}_{1,2}/{t}_3 )$, corresponding to the  other Dirac point. For the two solutions we find the frames,
\eq{
\label{eq:constrainedframe}
 \left(
 \begin{array}{cc}
 e_1^x & e_1^y \\
 e_2^x & e_2^y
 \end{array}
 \right) = -  \frac{1}{ \sqrt{3} T {t}_3} \, \mathbf{R}' \cdot \left(
 \begin{array}{cc}
 \pm 1 & 0 \\
0 & 1
 \end{array}
 \right) \cdot \left(
 \begin{array}{cc}
 \Delta & 0 \\
{t}_1^2 - {t}_2^2 & \sqrt{3} {t}_3^2
 \end{array}
 \right) 
 \\
}
where
$
 \mathbf{R}' = \left(
 \begin{array}{cc}
 \sin(\theta) &  \cos(\theta)  \\
- \cos(\theta) & \sin(\theta) 
 \end{array}
 \right) 
$
 is a frame rotation
 with $\theta = \phi + \partial_y {\Phi}$ and $\Delta$ is given by the two solutions of, 
\eqn{
\Delta^2 = (\sum_n t_n^2)^2 - 2 \sum_n t_n^4  \, .
}
The plus sign above corresponds to the solution which, for undeformed $t_n$, gives the $K$-point. The minus sign is for the solution giving the $K'$-point for undeformed $t_n$.
For both of these, the corresponding spatial metric that the Dirac field sees, which we will call the `electrometric' is,
\eq{
\label{eq:electrometric}
g_{ij} = e^a_i e^b_j \eta_{ab} =   \frac{ 3 T^2 }{   \Delta^2} \sum_n (\delta_{ij} - \frac{4}{3} \ell_n^i \ell_n^j )t_n^2 \, .
}
In the special case of equal $t_n = t(\bx)$, then the electrometric is Weyl flat, $g_{ij} = ( T/t(\bx) )^2 \delta_{ij}$. Recalling the strain gauge field vanishes,  this case  can then be
interpreted as massless Dirac
fields coupled only to a curved geometry and no
strain gauge field.

Strictly imposing the constraint~\eqref{eq:constraint} gives an exactly vanishing strain gauge field. However if we define the map from the lattice to the hopping functions with a subleading behaviour,
 \eqn{
 t_{n,\bx} &=&  t_n({\bx},a) = \bar{t}_n(\bx) + a \tau_n(\bx) + O(a^2)  
  }
  and for the phase we write,
  \eqn{
 \Phi({\bx},a) = \bar{\Phi}(\bx) + a \chi(\bx) + O(a^2)  
  } 
   and impose the constraint only on the $\bar{t}_n$ so it is not exactly satisfied, but is as $a \to 0$, then we may retain an $O(a^0)$ strain gauge field,
\begin{widetext}
\eq{
A_i  = \pm \frac{\Delta}{2\sqrt{3} \bar{t}_1 \bar{t}_2} 
   \left(
 \begin{array}{ccc}
- \frac{\bar{t}_1 + \bar{t}_2}{(\bar{t}_1+ \bar{t}_2)^2 - \bar{t}_3^2} 
&  \frac{\bar{t}_1 - \bar{t}_2}{(\bar{t}_1- \bar{t}_2)^2 - \bar{t}_3^2} 
&
\frac{4 \bar{t}_1 \bar{t}_2 \bar{t}_3}{\Delta^2}
\\
 \frac{1}{\sqrt{3}} \frac{\bar{t}_1 - \bar{t}_2}{(\bar{t}_1+ \bar{t}_2)^2 - \bar{t}_3^2} 
 & - \frac{1}{\sqrt{3}}  \frac{\bar{t}_1 + \bar{t}_2}{(\bar{t}_1- \bar{t}_2)^2 - \bar{t}_3^2}
& - \frac{4}{\sqrt{3}} \frac{(t_1^2-t_2^2) t_1 t_2}{\Delta^2 t_3}
 \end{array}
 \right)
   \left(
 \begin{array}{c}
\tau_+ \\
\tau_- \\
\tau_3
 \end{array}
 \right) - \partial_i \chi
}
\end{widetext}
where $\tau_\pm = \tau_1 \pm \tau_2$. The signs correspond to the solutions as for the frame in equation~\eqref{eq:constrainedframe}.
Taking $\Phi \to \Phi + a \,\delta \Phi$ then $A_i$ transforms by a gauge transformation, $A_i \to A_i - \partial_i \delta \Phi$ and the frame is invariant to leading order in $a$ as we require.

For general spatially varying hopping functions satisfying the constraint~\eqref{eq:constraint}
 we can define a wavevector $\vec{Q}(t_n(\vec{x}))$ by using exactly the same expressions as in equations~\eqref{eq:XY_defn} and~\eqref{eq:XY} for the homogeneous case, but now with varying $t_n(\vec{x})$. Explicit calculation of the magnetic field of $\vec{A}$ from its expression above shows it is simply related to the curl of this wavevector $\vec{Q}$ as,
\eq{
 \frac{1}{a} \left( \partial_x Q_y - \partial_y Q_x \right)=F_{xy} =\partial_x A_y - \partial_y A_x \; . 
}
Hence up to a gauge transformation we see that $\bQ/a$ is equal to the gauge field $\vec{A}$.
Finally we see that the constraint condition~\eqref{eq:constraint} which is that $\nabla \times \vec{Q}$ vanishes is in fact just the requirement that the strain
 magnetic field is indeed finite as $a \to 0$.

\section{Fine tuning and embedding}
\label{sec:Embedding}

Can this fine tuning of equation~\eqref{eq:constraint} arise from mechanical considerations?
Consider an almost flat embedding into $\mathbb{R}^3$ with coordinates $(X^i,Z)$ given by,
\eq{
X^i = x^i + \epsilon v^i(\vec{x}) \; , \quad Z = \sqrt{\epsilon} h(\vec{x})
}
with height function  $h$  and strain field $v^i$. Linearizing in $\epsilon$ the induced metric of this embedding is,
\eq{
\label{eq:induced}
g_{ij}^{(\mathrm{ind})} = \delta_{ij} + 2 \epsilon \sigma_{ij} \; , \quad 
\sigma_{ij} = \frac{1}{2} \left( \partial_i v_j + \partial_j v_i + \partial_i h \partial_j h \right)
}
with indices lowered/raised using $\delta_{ij}$ and $\sigma_{ij}$ is the strain tensor. Assuming the hopping parameters depend on bond length and not angle~\cite{manes2013generalized},
then to $O(\epsilon)$,
\eq{
t_n \simeq T \left( 1 - \epsilon \beta \sigma_{ij} \ell^i_n \ell^j_n \right)
}
where $\beta$ can be estimated for graphene as $\beta \simeq 3.3$~\cite{ribeiro2009strained}. 
We may neatly express our fine-tuning condition~\eqref{eq:constraint} as,
$
K^{ijk} \partial_k \sigma_{ij} = 0 
$.
Consider the canonical elastic energy,
\eq{
E_{\mathrm{mech}} =  \int d^2\vec{x} \left( \frac{\kappa}{2} (\partial^2 h)^2 + \mu \sigma_{ij}^2 + \frac{\lambda}{2}  \sigma_{ii}^2 \right)
}
with bending rigidity $\kappa$ and Lam\'e coefficients $\mu$ and $\lambda$. Varying the strain field $v^i$ yields, $\mu \partial_j \sigma_{ji} + \frac{\lambda}{2} \partial_i \sigma_{jj} = 0$. Assuming $\mu, \lambda > 0$ then this is generally incompatible with the previous fine tuning condition. Thus this membrane energetics does not impose the necessary constraint on strain for a curved space Dirac description.

In the event that the constraint is satisfied the perturbation to the electrometric is $g_{ij} \simeq \delta_{ij} + 2 \epsilon \beta \sigma_{ij}$.
As emphasized in \cite{yang2015dirac,stegmann2016current} this is not the same as the induced metric in equation~\eqref{eq:induced}. 
Non-perturbatively the $t_n({\bx})$ will be a functional of the induced metric $g_{ij}^{(\mathrm{ind})}({\bx})$, and the map to the electrometric is given by equation~\eqref{eq:electrometric}.

\section{Conclusion}
\label{sec:Conclusion}

We have argued that contrary to graphene folklore, the tight binding model with generic slow spatial variation of the hopping functions does not have a curved space Dirac description coupled to a strain gauge field. We find a continuum spinor description, but this is generally obstructed from truncating to first spinor derivatives by large magnetic fields. 
Making the $t_n$ vary perturbatively cannot solve this, although it does allow a consistent flat space  Dirac description with strain gauge field. 
However for generic slow non-perturbative variation of the $t_n$ there is no Dirac description at all. Related issues have been noted in the lattice literature when studying Euclidean theories on the honeycomb lattice as well \cite{Chakrabarti:2009sa}.
Standard examples of constant uniaxial strain, which have been well studied in the literature \cite{vozmediano:physrev}, do not in any way disagree with this result, as a constant strain can be interpreted as a constant metric deformation, which induces no curvature.

One may obtain a curved space Dirac description if one fine tunes the variation of the hopping functions. However this fine tuning appears unnatural and we have shown simple membrane energetics will not impose it.
Thus we believe that elastically deformed graphene monolayers do not generally have a curved space Dirac description. Likewise, optical lattice constructions of graphene-like lattices \cite{lee2009ultracold} will need to be highly fine-tuned to recover curved space Dirac.
This clearly has important implications for using graphene and other lattice systems as a laboratory to study curved space quantum field theory and analog gravity, as well as requiring a new paradigm to understand transport.

\subsection*{Acknowledgments}
This work was supported by STFC Consolidated Grant ST/T000791/1.

 \addcontentsline{toc}{section}{Bibliography}
\bibliography{Hofstadter}

\end{document}